\documentclass[twocolumn,fleqn,usenatbib]{mnras}
\usepackage{newtxtext,newtxmath}

\usepackage[T1]{fontenc}
\usepackage{ae,aecompl}

\usepackage{graphicx}	
\usepackage{amsmath}	
\usepackage{amssymb}	
\usepackage{xcolor}

\title[Observations of interstellar comet 2I/Borisov]{Visible and near-infrared observations of interstellar comet 2I/Borisov with the 10.4-m GTC and the 3.6-m TNG telescopes}

\author[J. de Le\'on et al.]{J. de Le\'on,$^{1,2}$\thanks{E-mail: jmlc@iac.es}
J. Licandro,$^{1,2}$
C. de la Fuente Marcos,$^{3}$
R. de la Fuente Marcos,$^{4}$
\newauthor
L. M. Lara,$^{5}$
F. Moreno,$^{5}$
N. Pinilla-Alonso,$^{6}$
M. Serra-Ricart,$^{1,2}$
M. De Pr\'a,$^{6}$
\newauthor
G.P. Tozzi,$^{7}$
A. C. Souza-Feliciano,$^{6,8}$
M. Popescu,$^{9}$
R. Scarpa,$^{10,1}$
J. Font Serra,$^{10,1}$
\newauthor
S. Geier,$^{10,1}$
V. Lorenzi,$^{11,1}$
A. Harutyunyan,$^{11}$
A. Cabrera-Lavers$^{10,1}$
\\
$^{1}$Instituto de Astrof\'{i}sica de Canarias (IAC), C/V\'ia L\'actea s/n, E-38205, La Laguna, Spain\\
$^{2}$Departamento de Astrof\'isica (ULL), E-38205, La Laguna, Spain\\
$^{3}$Universidad Complutense de Madrid, Ciudad Universitaria, E-28040 Madrid, Spain\\
$^{4}$AEGORA Research Group, Facultad de Ciencias Matem\'aticas, Universidad Complutense de Madrid, Ciudad Universitaria,\\
E-28040 Madrid, Spain\\
$^{5}$Instituto de Astrof\'isica de Andaluc\'ia - CSIC, Glorieta de la Astronom\'ia s/n, E-18008, Granada, Spain\\
$^{6}$Florida Space Institute, 12354 Research Parkway Partnership 1 Building, Suite 214 Orlando, FL 32826-0650, USA\\
$^{7}$Osservatorio Astrofisico di Arcetri, INAF, Lago E. Femi 6, I-50125, Firenze, Italy\\
$^{8}$Observat\'orio Nacional, Rio de Janeiro, 20921-400, Brazil\\
$^{9}$Astronomical Institute of the Romanian Academy, 5 Cu\k{t}itul de Argint, 040557 Bucharest, Romania\\
$^{10}$GRANTECAN, Cuesta de San Jos\'e s/n, E-38712 Bre\~{n}a Baja, La Palma, Spain\\
$^{11}$Fundaci\'on Galileo Galilei - INAF, Rambla Jos\'e Ana Fern\'andez P\'erez 7, E-38712 Bre\~{n}a Baja, La Palma, Spain\\
}

\date{Accepted XXX. Received YYY; in original form ZZZ}

\pubyear{2020}

\begin{document}
\label{firstpage}
\pagerange{\pageref{firstpage}--\pageref{lastpage}}
\maketitle

\begin{abstract}
In this work, we present the results of an observational study of 2I/Borisov carried out with the 10.4-m Gran Telescopio Canarias (GTC) and the 3.6-m Telescopio Nazionale Galileo (TNG), both telescopes located at the Roque de Los Muchachos Observatory, in the island of La Palma (Spain). The study includes images in the visible and near-infrared, as well as visible spectra in the 3600 -- 9200 \AA \ wavelength range. $N$-body simulations were also performed to explore its orbital evolution and Galactic kinematic context. The comet's dust continuum and near-infrared colours are compatible with those observed for Solar system comets. From its visible spectrum on the nights of 2019, September 24 and 26 we measured CN gas production rates $Q$(CN) = (2.3 $\pm$ 0.4) $\times$ 10$^{24}$ mol s$^{-1}$ and $Q$(CN) = (9.5 $\pm$ 0.2) $\times$ 10$^{24}$ mol s$^{-1}$, respectively, in agreement with measurements reported by other authors on similar nights. We also obtained an upper limit for the C$_2$ production rate of $Q$(C$_2$) $<$ (4.5 $\pm$ 0.1) $\times$ 10$^{24}$ mol s$^{-1}$. Dust modelling results indicate a moderate dust production rate of $\sim$50 kg s$^{-1}$ at heliocentric distance $r_h$=2.6 au, with a differential power-law dust size distribution of index $\sim$ --3.4, within the range reported for many comet comae. Our simulations show that the Galactic velocity of 2I/Borisov matches well that of known stars in the solar neighbourhood and also those of more distant regions of the Galactic disc.
\end{abstract}




\section{Introduction}

Until recently, the lack of detections of interstellar comets posed a problem to the theories of planet and comet formation (see e.g. \citealt{mcglynn1989}). The discovery rate of such comets could be used to estimate the fraction of stars with planets \citep{stern1990} and the absence of detections could be regarded as natural \citep{sen1993}, but also as resulting from technical limitations \citep{engelhardt2014,cook2016}. The discovery of the first interstellar minor body, 1I/2017~U1 (`Oumuamua), that was rather small and lacked clear signs of cometary activity (see e.g. \citealt{bannister2019}) could neither provide a conclusive answer to the issue of how numerous interstellar comets are, nor fully support standard theories of planet and comet formation. The discovery and study of additional interstellar minor bodies may provide us with the framework required to understand whether and to what extent `Oumuamua is a typical member of this dynamical class or perhaps an outlier.

     \begin{table*}
      \centering
        \caption{Heliocentric and barycentric orbital elements and 1 $\sigma$ uncertainties of interstellar comet 2I/Borisov. This solution is hyperbolic with a statistical significance of 155164 $\sigma$ (barycentric) and it is based on 1310 observations that span a data-arc of 444 days. The orbit determination has been computed by D. Farnocchia at epoch JD 2459061.5 that corresponds to 00:00:00.000 TDB, Barycentric Dynamical Time, on 2020, July 31, J2000.0 ecliptic and equinox. Source: JPL's Solar System Dynamics Group Small-Body Database (SSDG SBDB; solution date, 2020-Mar-19 08:23:45).}\label{elements}
        \begin{tabular}{lcc}
           \hline
            Orbital parameter                                    & Heliocentric       & Barycentric \\
           \hline
            Perihelion distance, $q$ (au)                  & 2.006624 $\pm$ 0.000002               &    2.011869 \\
            Eccentricity, $e$                                      & 3.35619 $\pm$ 0.00002                 &    3.35881  \\
            Inclination, $i$ (\degr)                                                 & 44.052626 $\pm$ 0.000011              &   44.062226 \\
            Longitude of the ascending node, $\Omega$ (\degr)                       &  308.14892 $\pm$ 0.00003                 &  308.10039  \\
            Argument of perihelion, $\omega$ (\degr)                                &  209.12461 $\pm$ 0.00005                 &  209.16747  \\
            Mean anomaly, $M$ (\degr)                                                & 295.277 $\pm$ 0.003                     &  294.539    \\
            \hline
            Non-gravitational radial acceleration parameter, $A_1$ (au d$^{-2}$)    &  \multicolumn{2}{c}{7.31 $\times$ 10$^{-8}$ $\pm$ 4.2 $\times$ 10$^{-9}$}            \\ 
            Non-gravitational transverse acceleration parameter, $A_2$ (au d$^{-2}$) &  \multicolumn{2}{c}{--3.3 $\times$ 10$^{-8}$ $\pm$ 1.1 $\times$ 10$^{-8}$}             \\ 
           \hline           
        \end{tabular}
        \end{table*}

Comet C/2019 Q4 was discovered as gb00234 by G. Borisov on 2019, August~30, observing from MARGO (Mobile Astronomical Robotics Genon Observatory), Nauchnij, in the Crimean peninsula.\footnote{\url{https://minorplanetcenter.net/mpec/K19/K19RA6.html}} It was found at a solar elongation of just 38\degr\,when the object was moving inbound at about 3 au from the Sun, and it was soon identified as having a hyperbolic orbit, being officially named as 2I/Borisov by the IAU on 2019, September 24:\footnote{\url{https://minorplanetcenter.net/mpec/K19/K19S72.html}} this is only the second interstellar object known. Table~\ref{elements} shows the latest orbit determination of 2I/Borisov (as of 2020, April 22), based on 1310 observations that span a data-arc of 444 days. Its current path is hyperbolic with a statistical significance above 155164 $\sigma$. Contrary to what happened with 1I/`Oumuamua, comet 2I/Borisov was discovered when it was entering the Solar system, and so it is observable for at least one year from its discovery. First results from different works showed that it presents visible colours, gas and dust production rates, and nuclear properties that are similar to those observed for Solar system comets \citep{fitzsimmons2019,jewitt2019,opitom2019,guzik2020,jewitt2020,kareta2020,mckay2020}. Here, we investigate observationally the cometary activity of 2I/Borisov using images in the visible and the near-infrared, and its spectral properties in the visible region using low-resolution spectroscopy. In addition, we use its orbit determination to explore numerically its dynamical evolution, aiming at placing it within its Galactic context. This paper is organised as follows. In Sect. \ref{obs}, we describe observations and data reduction. In Sect. \ref{results}, we analyse the observed coma and the spectral properties of 2I/Borisov, and compare these observational results with those of Solar system comets and other related populations. In addition, we explore the pre- and post-encounter orbital evolution of 2I/Borisov as well as investigate its Galactic context. Our conclusions are laid out in Sect. \ref{conclusions}.

\section{Observations and Data Reduction}
\label{obs}

We obtained low-resolution visible spectra and images of 2I/Borisov on 2019, September 13, 24, and 26, using the Optical System for Imaging and Low Resolution Integrated Spectroscopy (OSIRIS) camera-spectrograph \citep{cepa2000,cepa2010}, at the 10.4-m Gran Telescopio Canarias (GTC), and near-infrared images on 2019, September 24 using the NICS camera-spectrograph \citep{baffa2001} at the 3.6-m Telescopio Nazionale Galileo (TNG). Both telescopes are located at the Roque de Los Muchachos Observatory (ORM), in the island of La Palma (Canary Islands, Spain). Observational details are shown in Table \ref{observations}.

\subsection{Visible and near-infrared images}

\begin{figure*}
    \centering
	\includegraphics[width=1.6\columnwidth]{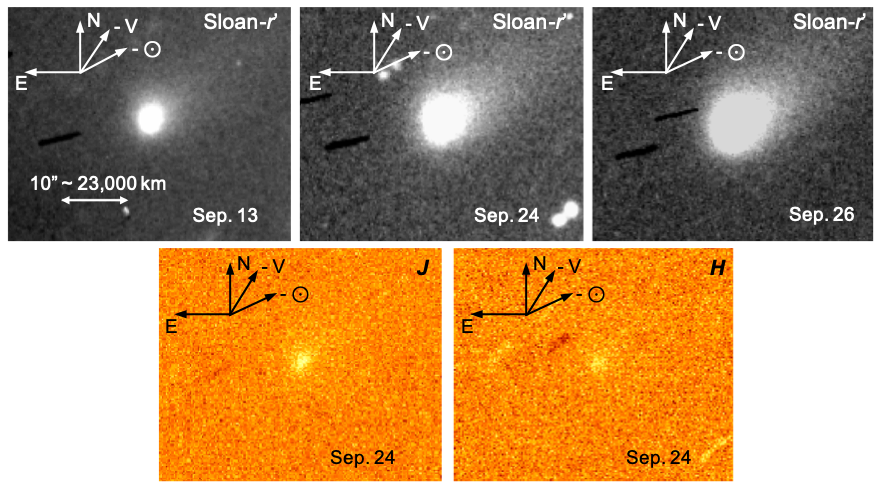}
    \caption{Images of 2I/Borisov obtained on 2019, September 13, 24, and 26 using the OSIRIS instrument and the Sloan $r'$ filter at the GTC (upper panels). The first image is a composite of four individual images, while the second and the third are composites of two individual images each. The bottom panels show the images obtained on September 24 using the NICS instrument and the $J$ and $H$ filters at the TNG. Both images are composite of 10 individual images each. Arrows show the directions of north (N) and east (E), as well as the projected anti-Solar vector (-$\sun$) and the negative of the orbital velocity vector (-V) on each night. The spatial scale is also included on the first image and it is the same for all of them.}
    \label{figure1}
\end{figure*}

The OSIRIS detector is a mosaic of two Marconi 2048$\times$4096 pixel CCDs. The total unvignetted field of view is 7.8$\times$7.8 arcminutes, and the plate scale is 0.127~"/pix. Standard operation mode consists of a 2$\times$2 binning, with a readout speed of 200~kHz (with a gain of 0.95~e-/ADU and a readout noise of 4.5~e-). On the 3 nights of 2019 September we obtained individual images using the Sloan {\em r'} filter and an exposure time of 30 seconds, with the telescope tracking at the comet's proper motion. The target was visible only at low elevation (about 25$\degr$\,above local horizon) and during the twilight, therefore observations were extremely challenging. Four images were obtained on the night of September 13, at an airmass ranging from 2.53 to 2.31. On the nights of September 24 and 26, we acquired two images each night, with an airmass of 1.91 and 1.98, respectively. Photometric data on the three nights were reduced using standard tasks in IRAF, following a procedure similar to the one described in \citet{licandro2019}. Images in the Sloan {\em r'} filter were bias and flat-field corrected (using sky flats), and aperture photometry was computed using a 5" aperture in all cases, which was equivalent to a projected radius of $\sim$12,000 km on the night of Sep. 13, and $\sim$11,000 km on the nights of Sep. 24 and Sep. 26. Background sky was measured (and subsequently subtracted) in a concentric annulus extending from 10" to 15". Flux calibration was done using GTC zero-points computed for each observing night and provided by the telescope support astronomer. The images of each observing night were aligned on the comet opto-center and averaged (see Fig.~\ref{figure1}).

Broad-band photometry in the near-infrared was performed on the night of September 24 using the NICS camera-spectrograph. The plate scale was 0.25"/pix, yielding a field of view of 4.2$\times$4.2 arcminutes. The series of images used the standard Johnson $J$, and $H$ filters and consisted of 10 individual exposures of 60 seconds following a dithering pattern on different positions on the CCD, separated by off-sets of 10 pixels. The tracking of the telescope was set at the proper motion of the target. The comet was observed at an airmass of 2.22 for the $J$ filter and 2.02 for the $H$ filter. The data were reduced in the standard way using IRAF routines and all frames were flat-field corrected and sky subtracted. Standard aperture photometry was done, using an aperture of 5". We observed one field of standard stars for calibration, AS04-0\_FS04 \citep{persson1998}, using a mosaic of five individual exposures of 5 seconds. The combined images in the $J$ and $H$ filters are also shown in Fig. ~\ref{figure1}.

\subsection{Visible spectra}
\label{vis}

We obtained two different sets of visible spectra of comet 2I/Borisov with OSIRIS at the GTC. The first set, on the night of September 13, consisted of three individual spectra of 300 seconds exposure each, using the R300R grism, in combination with a second-order spectral filter. This produced a spectrum in the range 4900 to 9200 \AA, with a dispersion of 7.74 \AA/pix for a 0.6" slit width. We used a 1.23" slit, oriented in the parallactic angle in order to minimise losses due to atmospheric dispersion, and the telescope tracking was at the comet's proper motion. We offset the telescope 10" between spectra in the slit direction to better correct for fringing and increase the signal-to-noise ratio (SNR). To correct for telluric absorptions and obtain the relative reflectance spectrum of the comet, we also observed three G2V stars -- SA93-101, SA98-978 and SA115-271 -- from the Landolt catalogue \citep{landolt1992}, immediately before observing the object, and at similar airmass using the same spectral configuration. The spectrum of the comet was divided by those of the solar analogue stars, and the resulting spectra were finally averaged. Preliminary results of the data acquired on Sep. 13 were presented in \citet{deleon2019}. The second set, obtained on the nights of September 24 and 26, consisted of another three 300 seconds individual spectra on each night, but this time using the R300B grism, covering a wavelength range from 3600 to 7500 \AA, and with a dispersion of 4.96\AA/pix for a 0.6" slit. We used 2.52" and 1.23" slit widths on Sep. 24 and Sep. 26, respectively, oriented in the parallactic angle. We also observed the solar analogue star SA98-978 from the Landolt catalogue to obtain the reflectance spectra on both nights, using the same procedure described above. The two reflectance spectra (R300R and R300B) are shown in Fig. \ref{spectrum}. To compute gas production rates of different cometary species, we flux calibrated the spectra obtained with the R300B grism on the nights of Sep. 24 and Sep. 26. We only used two out of the three individual spectra obtained on each night, discarding the first individual spectra as it presented a lower SNR. In both cases, we used the spectrophotometric standard star G191-B2B.

%
%

     \begin{table}
      \centering
        \caption{Observational circumstances of the data presented in this work, obtained in 2019 September. Information includes date, telescope (Tel.), airmass (X), heliocentric ($r_h$) and geocentric ($\Delta$) distances, phase angle ($\alpha$), position angle of the projected anti-Solar direction ($\theta_{\sun}$) an the position angle of the projected negative heliocentric velocity vector ($\theta_{-V} $). Orbital values have been taken from JPL's HORIZONS system.}\label{observations}
        \begin{tabular}{lccccccc}
           \hline
            Date & Tel. & X & $r_h$  & $\Delta$ & $\alpha$  & $\theta_{\sun}$ & $\theta_{-V}$\\
                             &           &       & (au) & (au)        & ($^{\circ}$) & ($^{\circ}$) & ($^{\circ}$) \\
           \hline
           13.24 & GTC & 2.42& 2.766 & 3.406 & 14.5 & 298.6 & 326.8\\
           24.25 & GTC & 1.91 & 2.607 & 3.165 & 16.7 & 295.9 & 328.2\\
                     & TNG & 2.12     &            &           &         &            &         \\
           26.24 & GTC & 1.98 & 2.579 & 3.122 & 17.1 &295.5 & 328.4 \\
           \hline
        \end{tabular}
        \end{table}

\section{Results and Analysis}
\label{results}

\subsection{Nuclear size}
\label{nucleus}

Images of comet 2I/Borisov in the visible showed a conspicuous comet-like coma and tail, as it can be seen in the upper panels of Fig.~\ref{figure1}. The images in the near-infrared have a lower SNR but still some cometary shape is marginally detected (Fig.~\ref{figure1}, lower panels). Apparent magnitudes $r'$ = 18.23 $\pm$ 0.04,  $r'$ = 17.68 $\pm$ 0.02, and $r'$ = 17.62 $\pm$ 0.02 were derived for the nights of 2019, Sep. 13, 24, and 26, respectively. From the apparent magnitudes, we derived the absolute magnitudes using Eq. (1) from \citet{jewitt2019}, obtaining $H_r = 12.72\pm0.05$, $H_r = 12.43\pm0.05$, and $H_r = 12.41\pm0.05$ for the nights of Sep. 13, 24, and 26, respectively. Assuming that all the light within the aperture of 5" used to do the photometry is actually coming from scattered light from the nucleus, and using an albedo of $p_V = 0.04$ (typical of comet nuclei, see \citealt{licandro2018}), the computed absolute magnitudes correspond to a nuclear radius of $\sim$ $3.6$~km. However, as the comet is active, its brightness also depends on the contribution of dust in the photometric aperture so we consider this an upper limit. This value is in agreement with the range of nuclear sizes reported by other authors also from photometry: \citet{jewitt2019} set a range between 0.35 and 3.8 km, assuming an albedo of $p_V$ = 0.1, while \citet{fitzsimmons2019} constrained the nuclear radius to 0.7-3.3 km. The size is compatible to the typical sizes observed for cometary nuclei in our Solar system ($\sim$ 2.8 km, \citealt{meech2017}). From their observation in the $K$-band using the Gemini South telescope, \citet{lee2019} reported a nuclear size of $r_N$ = 1.5 km, assuming an infrared albedo of 0.07. A more recent work from \citet{jewitt2020} using observations of 2I/Borisov made with the Hubble Space Telescope has constrained the size of the nucleus between 0.2 and 0.5 km. This size is similar to the size of the other only known interstellar object, 1I/`Oumuamua ($\lesssim$ 200 m), that showed none or very weak activity \citep{micheli2018}. A recent work on the size-frequency distribution of long-period comets detected by the Pan-STARRS1 near-Earth object survey, has identified a lack of objects with diameters $<$ 1km and showing activity \citep{boe2019}. There is no clear explanation on why small Oort cloud objects might not be active and behave differently from large ones. On the contrary, we find both active and non-active small objects among the current interstellar sample. Nevertheless, this sample is extremely limited and needs to be significantly enlarged in order to properly compare both populations.

\subsection{Dust colours}
\label{colours}

\begin{figure}
    \centering
     \includegraphics[width=1\columnwidth]{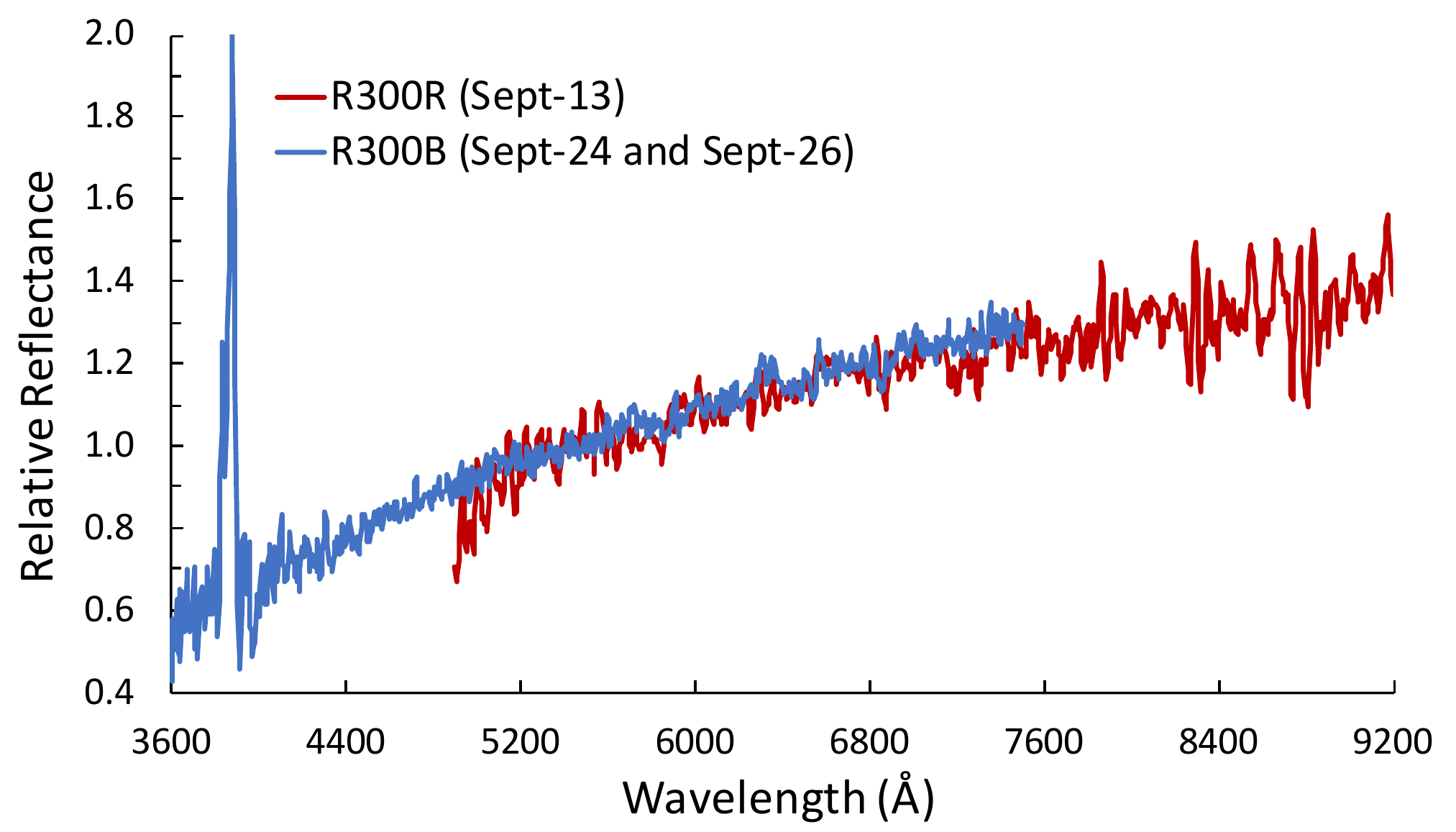}
     \caption{Visible reflectance spectrum of comet 2I/Borisov obtained with GTC. The R300R spectrum (red) was obtained on the night of 2019, Sep. 13, and presented as a preliminary measurement in \citet{deleon2019}. The R300B one (blue) is the result of averaging the spectra obtained on the nights of 2019, Sep. 24 and 26, using also GTC.  The two spectra are normalised to unity at 5500 \AA.}
     \label{spectrum}
   \end{figure}

The two individual spectra of 2I/Borisov obtained with the R300B (blue) and R300R (red) grisms are shown in Fig. \ref{spectrum}, and match perfectly in the common wavelength interval, i.e., from 4900 to 7500 \AA. We used this interval to join the two spectra. We computed the spectral slope $S'$ using this composite spectrum in the range 4000 -- 9000 \AA, following the $S'$ definition in \citet{luu1996}. The obtained value was $S' = 12 \pm 1$ \%/1000\AA. The quoted uncertainty in the value of $S'$ has been computed as the standard deviation ($\sigma$) of the $S'$ values obtained for each single reflectance spectrum of the object as in \citet{licandro2019}. The computed spectral slope of comet 2I/Borisov fits well into the general spectral behaviour of cometary dust, having spectral gradients similar to those found for X- and D-type asteroids \citep{jewitt2015,licandro2018}, and it is also similar to the computed spectral slope for 1I/`Oumuamua in the same wavelength range, $S'$ = 10 $\pm$ 6 \%/1000\AA \ \citep{ye2017}. Our computed value of the spectral slope $S' = 22 \pm 1$ \%/1000\AA \ in the range 3900 -- 6000 \AA, supports the results obtained by  \citet{fitzsimmons2019}, where they found a value of $19.9 \pm 1.5$ \%/1000 \AA \ in the same wavelength interval measured on a visible spectrum obtained on the night of 2019, Sep. 20, and reinforces the conclusion that comet dust from 2I/Borisov behaves likes normal comet dust observed in our Solar system, presenting a steeper slope at shorter wavelengths. 

We also obtained the ($g - r'$) colour from our visible reflectance spectrum. To do this, we obtained the reflectances $R_g$ and $R_{r'}$ through the convolution of the Sloan transmission curves of these two filters and the visible spectrum. Then, we transformed such reflectances into colours using the expression ($g - r'$) = --2.5 log($R_g$/$R_{r'}$) + ($g - r'$)$_{\sun}$, where ($g - r'$)$_{\sun}$ is the colour index of the Sun, taken to be equal to 0.45 from \citet{holmberg2006}. Using this expression, we get a value of ($g - r'$) = 0.69 $\pm$ 0.04, in agreement within the errors with the value of ($g - r'$) = 0.66 $\pm$ 0.01 reported by \citet{guzik2020}. \cite{jewitt2019} reported a colour of ($B-V$) =  0.80 $\pm$ 0.05. This value translates into a value of ($g - r'$) = 0.60 $\pm$ 0.04, using the transformation discussed by \citet{jester2005}, which is slightly smaller than the ($g - r'$) colour computed in this work and in \citet{guzik2020}. From the images of comet 2I/Borisov in the near-infrared on the night of Sep. 24, we obtained $J$ =  16.97 $\pm$ 0.31 and $H$ = 16.45 $\pm$ 0.44, which gives ($J - H$) = 0.52 $\pm$ 0.54. 

\begin{figure}
    \centering
     \includegraphics[width=0.9\columnwidth]{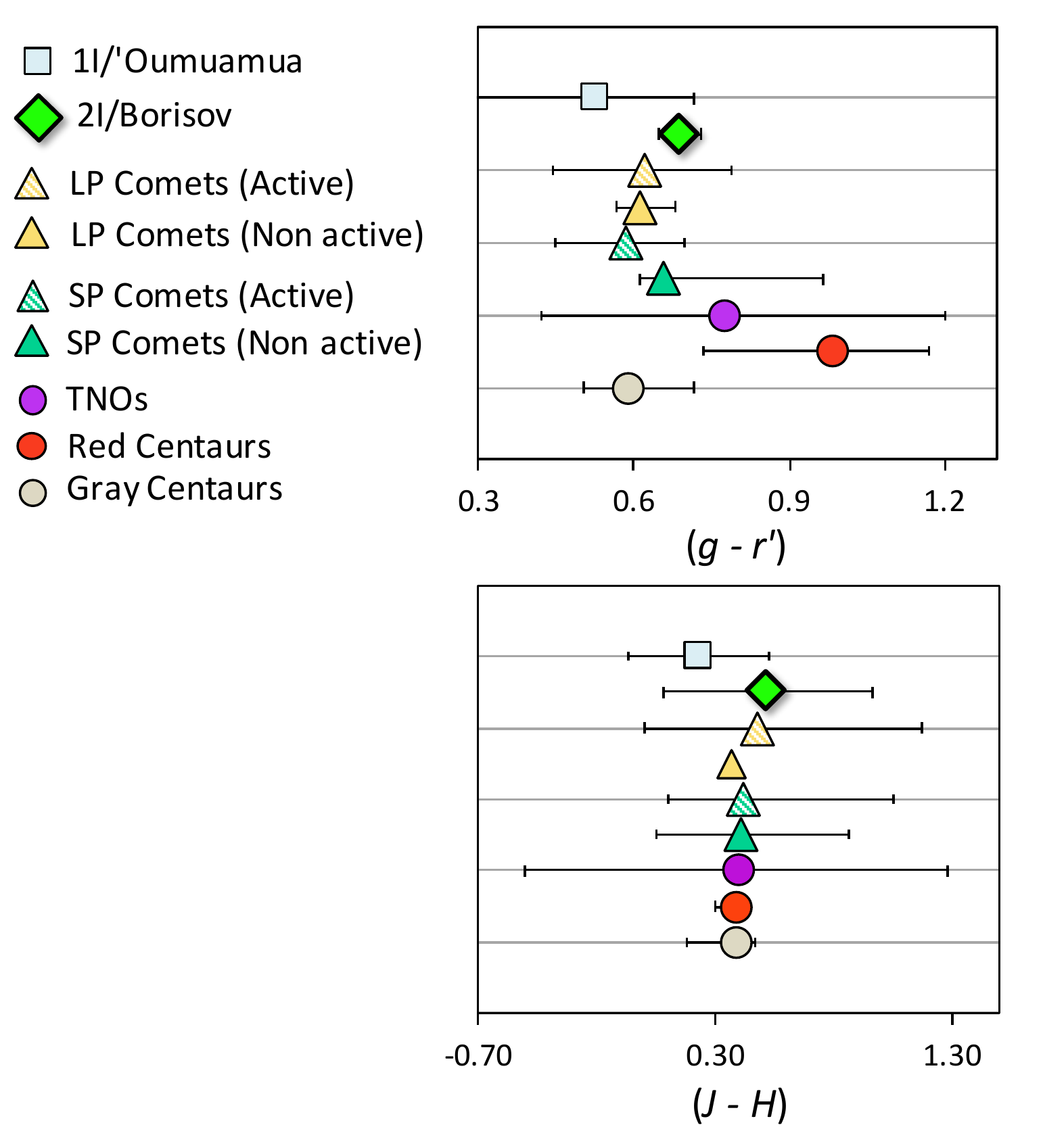}
     \caption{Comparison of the ($g-r'$) and ($J-H$) colours of comet 2I/Borisov (this work) to the colours of long period comets (LP), short period comets (SP), TNOs, and red and gray Centaurs. For these populations, we plot the median value of the population as well as the maximum and minimum values (error bars). In the case of 2I/Borisov and  the ($J-H$) colour of 1I/`Oumuamua, error bars correspond to the error associated with each computed colour. See Table \ref{infocolors} and main text for more details.}
     \label{colors}
   \end{figure}

We now compare the obtained colours for comet 2I/Borisov to those of the only other interstellar object observed so far, 1I/`Oumuamua, and with the colours of comets and other related populations in our Solar system such as trans-Neptunian objects (TNOs) and Centaurs. Fig. \ref{colors} shows the ($g-r'$) and ($J-H$) median values of these populations, as well as their maximum and minimum values as horizontal error bars. Following \citet{jewitt2015}, we distinguish between active and non-active comets, and between short-period (SP, i.e., Jupiter family comets) and long-period (LP) comets (including Halley type). These colours have been compiled from the literature using different sources. Table \ref{infocolors} shows the median value for each population, as well as the number of observations used to compute these median values ($N$) and the corresponding reference. We converted ($B-V$) colours from \citet{hainaut2012} into ($g-r'$) using the transformations presented in \citet{jester2005} and retaining only those values with an uncertainty ($B-V$) $<$ 0.15. In the same way, we only selected ($J-H$) values with an uncertainty $<$ 0.4. Finally, we used the ($g-r'$) colours of 1I/`Oumuamua from \citet{bannister2017}, \citet{jewitt2017}, \citet{ye2017}, and \citet{bolin2018}, while its ($J-H$) colour was obtained from the reflectance spectrum presented by \citet{fitzsimmons2018}, using the same procedure described above to obtain our ($g-r'$) colour in the case of 2I/Borisov. Both values are shown in Table \ref{infocolors} and, when compared to the median value of the different populations shown in Fig. \ref{colors}, it can be seen that comet 2I/Borisov presents visible and near-infrared colours that are slightly redder than those for 1I/`Oumuamua, but in good agreement with the colours of comets in the visible and comets, TNOs and Centaurs in the near-infrared. 

     \begin{table*}
      \centering
        \caption{Visible and near-infrared colours of comets and other related populations (TNOs and Centaurs) of our Solar system. We show the number of observations used ($N$) and the median value of each population, as well as the corresponding reference. We also include the colours of 2I/Borisov from this work and the literature, and those of 1I/`Oumuamua from the literature.}\label{infocolors}
        \begin{tabular}{lcccccc} 
           \hline
            \multicolumn{4}{c}{Visible} & \multicolumn{3}{c}{Near-infrared}\\
           \hline
                     & $N$ & ($g-r'$) & Ref. & $N$ & ($J-H$) & Ref.\\
           \hline
           Gray Centaurs & 15 & 0.59 &\citet{hainaut2012}  & 11 & 0.39 & \citet{hainaut2012}  \\
           Red Centaurs & 11 & 0.98 & \citet{hainaut2012} & 6 & 0.39 & \citet{hainaut2012} \\
           TNOs & 174 & 0.77 & \citet{hainaut2012} & 117 & 0.40 & \citet{hainaut2012} \\
           SP comets (active) & 22& 0.59 & \citet{solontoi2012} & 12 & 0.42 & \citet{hanner1984} \\
            & & & & & & \citet{popescu2016}\\
           SP comets (non-active) & 48 & 0.66 &\citet{lamy2009} & 2 & 0.41 & \citet{hainaut2012} \\
           & & & & & & \citet{sykes2000}\\
           LP comets (active) & 40 & 0.62 & \citet{jewitt2015} & 3 & 0.48 &\citet{picazzio2010}\\
            & & & \citet{solontoi2012} & & & \\
           LP comets (non-active) & 5 & 0.61 & \citet{lamy2009} & 1 & 0.37 & \citet{sykes2000} \\
           \hline
           1I/`Oumuamua & 4 & 0.53  & \citet{bannister2017} & \multicolumn{2}{c}{0.23 $\pm$ 0.25} & \citet{fitzsimmons2018} \\
                                    &   &          & \citet{jewitt2017}  &  &   & \\
                                    &   &          & \citet{ye2017}   &    &    & \\
                                    &   &          & \citet{bolin2018} &   &   & \\
           2I/Borisov & \multicolumn{2}{c}{0.69 $\pm$ 0.04} & This work & \multicolumn{2}{c}{0.52 $\pm$ 0.54} & This work\\
                            & \multicolumn{2}{c}{0.66 $\pm$ 0.01} & \citet{guzik2020} & & & \\
                            & \multicolumn{2}{c}{0.60 $\pm$ 0.04} & \citet{jewitt2019} & & & \\
           \hline
        \end{tabular}
        \end{table*}

\subsection{Dust tail model}
\label{model}

For the dust tail fitting, we used our Monte Carlo dust tail code that has been described in several works in the past to characterise the dust environments of comets and main-belt comets (see e.g. \citealt{moreno2016,moreno2017}). The particles are assumed to be spherical, and their trajectories are described by the $\beta$ parameter, defined as $\beta$ = ($C_{pr}Q_{pr}$)/($\rho$$d$), where $\rho$ is the particle density, $d$ is the particle diameter, $C_{pr}$=1.19$\times$10$^{-4}$ g cm$^{-2}$ is the radiation pressure coefficient, and $Q_{pr}\sim$1 is the scattering efficiency for radiation pressure. The code computes the position on the sky plane of particles ejected isotropically from a sublimating nucleus. Their trajectories depend on $\beta$ and their terminal velocities. The comet activity is assumed to start at a heliocentric distance of 4.5 au (2019, June 5), as estimated by \citet{jewitt2019}. Although the onset of the activity might have occurred earlier (at around 7.8 au, as suggested by \citealt{ye2020}), the dust production rate was likely very low at such distances so as to have a detectable influence on the modelling results. The particle density is assumed at $\rho$=800 kg m$^{-3}$, in line with Rosetta/GIADA estimates for comet 67P by \citet{fulle2016}, and the geometric albedo is set to 4 per cent, consistently with the value assumed in Section \ref{nucleus}. A phase function correction is performed by assuming a linear phase coefficient of $b$=0.03 mag deg$^{-1}$. The size distribution is assumed to be governed by a power-law function with power index $\kappa$. A broad size distribution is assumed, having minimum and maximum radii given by r$_{min}$=1 $\mu$m  and r$_{max}$=1 cm. For simplicity, the terminal speeds $v$ are parameterised by an expression of the type $v=v_0\beta^{\gamma}r_h^{-0.5}$, where $v_0$ is a time-independent speed parameter and $r_h$ is the comet heliocentric distance expressed in au. For an expanding gas flow from ice sublimation, $\gamma\approx$ 0.5. However, this parameter has been found to vary between 0.42 and 1.5 from Rosetta/GIADA measurements on comet 67P \citep{dellacorte2015,dellacorte2016}. On the other hand, measurements of individual particles from Rosetta/OSIRIS at far heliocentric distances by \citet{rotundi2015} show no dependence of particle speeds on size. Thus, we left $\gamma$ as a free parameter of the model. The mass loss rate is assumed to be related to the comet heliocentric distance by a function of the form $\log{dM/dt}=A\log{r_h}+B$ where $dM/dt$ is expressed in kg s$^{-1}$ and $r_h$ in au, and constants $A$ and $B$ are to be found from the model fittings. These two constants, together with $v_0$, $\gamma$, and $\kappa$ constitute the set of input parameters of the model. The best-fit parameters are determined by a minimisation routine \citep{nelder1965}.

   \begin{table}
      \caption{Summary of best-fit model parameters obtained after applying our Monte Carlo dust tail code to the images of comet 2I/Borisov acquired with GTC. See text for more details.}
         \label{Modpartab}
\centering                          
\begin{tabular}{c c c c c c}        
\hline\hline                 
$A$ & $B$ & $v_0$  (m s$^{-1}$)   & $\gamma$ & $\kappa$ \\
\hline                        
--4.92 & 3.77 &  54.6 & 0.1 & --3.46 \\
--4.79 & 3.61 & 152.9 & 0.5 & --3.44 \\
\hline                        
\end{tabular}
\end{table}

\begin{figure*}
   \centering
   \includegraphics[width=1.3\columnwidth]{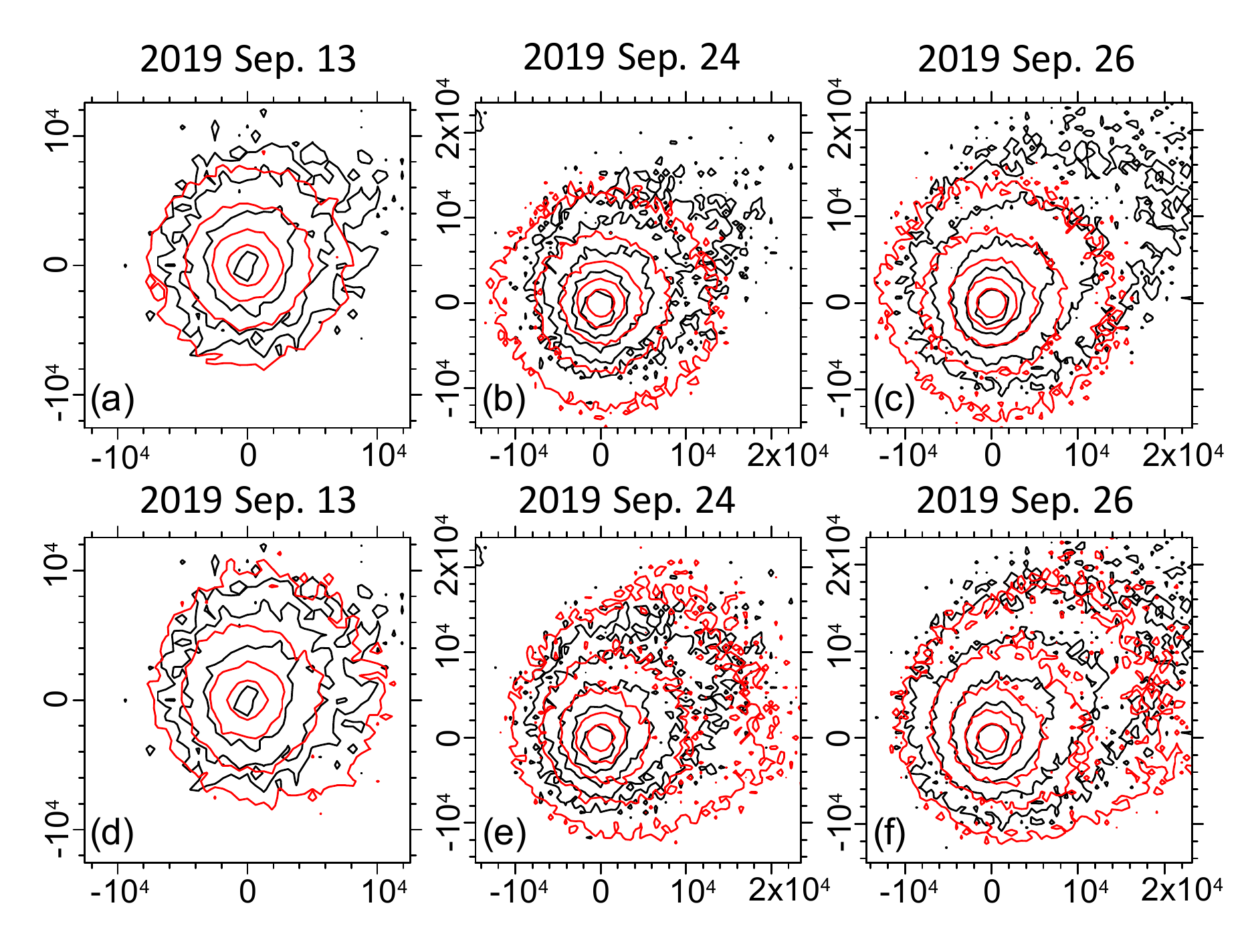}
   \caption{Observed (black contours) and computed (red contours) isophotes of the GTC images of the dust tail of comet 2I/Borisov. The images are all oriented North up, and East to the left. All axes scales indicate cometocentric distances in km. The uppermost three panels (a), (b), and (c) correspond to the case $\gamma$=0.5. Panels (d), (e), and (f) correspond to the model with $\gamma$ taken as a free parameter, resulting in $\gamma$=0.1 (see Table~\ref{Modpartab}). The innermost isophote on the 2019-Sep-13 image is 1.5$\times$10$^{-13}$ solar disk intensity units (sdu), and for the other two dates it is 2.0$\times$10$^{-13}$ sdu. Isophote levels decrease in factors of two outwards.}
     \label{Modelfig}
\end{figure*}

We arrived to the best-fit parameters shown in Table~\ref{Modpartab}, first row. In Fig.~\ref{Modelfig}, panels (d), (e), and (f), we show a comparison between the observed and modelled isophote fields. The resulting $A$ and $B$ parameters provide a dust production rate of 3.6 kg s$^{-1}$ at 4.5 au, increasing up to 52 kg s$^{-1}$ at 2.59 au,
 corresponding to the date of the latest observation. The value of the power index, $\kappa$=--3.4, is within the typical range of measured or inferred values for other
comets \citep{fulle2004}, and in the range estimated by \citet{guzik2020} for this object (--3.7 $\pm$ 1.8). For the derived size distribution parameters, the effective radius \citep{hansen1974} is r$_{eff}$=170 $\mu$m, close to the 100 $\mu$m value estimated by \citet{jewitt2019}. The coefficient $\gamma$=0.1 indicates a very weak dependence of terminal speeds on size. This is in line with the single particle detections speeds for 67P at large $r_h$ \citep{rotundi2015}. To show the results for $\gamma$=0.5, as expected from gas drag models, we run the code by fixing that parameter, and the results are also shown in Table~\ref{Modpartab} (second row) and Fig.~\ref{Modelfig}, panels (a), (b), and (c). The fittings are worser in this case, producing a more circularised isophote field than observed, and diverging from the observed isophotes at large nucleocentric distances. However, both models show consistently similar dust mass loss rates and size distribution (Table~\ref{Modpartab}). The corresponding particle speeds for $\beta$=1 are 94 m s$^{-1}$ (for $\gamma$=0.5) and 34 m s$^{-1}$ (for $\gamma$=0.1), which compare well with the estimated speed of 44 $\pm$ 14 m s$^{-1}$ by \citet{guzik2020} and speed range from 69 to 92 m s$^{-1}$ by \citet{kochergin2019}.

In all the simulations performed, the observed brightness is assumed to be dominated by the dust, i.e., the brightness of the nucleus is neglected. Model runs adding a nucleus at the opto-center indicate that the results are very similar for nucleus sizes of up to $\sim$3 km in radius, provided that the nucleus has the same geometric albedo and phase coefficient values as those assumed for the dust particles. 

\subsection{Gas production}
\label{gas}

\begin{figure}
   \centering
   \includegraphics[width=1\columnwidth]{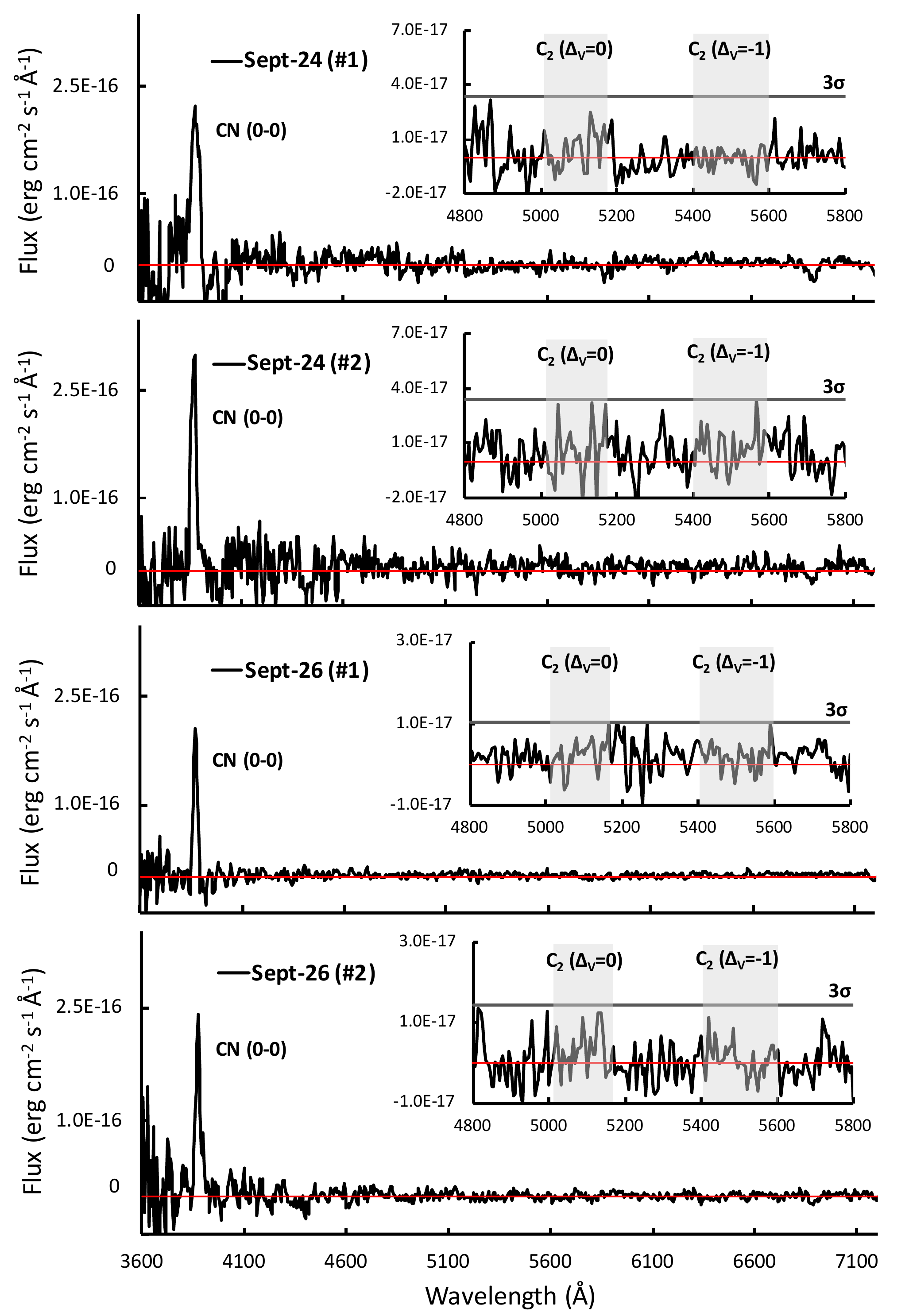}
   \caption{Emission spectra of interstellar comet 2I/Borisov obtained on the nights of 2019, September 24 and 26 using OSIRIS at the GTC. The CN emission band is clearly seen in all individual spectra.}
     \label{emission}
\end{figure}

To analyse the gas emission of comet 2I/Borisov, we studied the 2D flux-calibrated spectra described in Section \ref{vis}. For a better visualisation of the emission bands, we used a spectrum of the Sun downloaded from the CALSPEC compilation \citep{bohlin2014} to remove the continuum on each individual spectra: the CN (0-0) emission at 3880 \AA \ is clearly detected (see Fig. \ref{emission}), and we do not detect any C$_2$ emission within the 3 $\sigma$ level. For the 2D analysis we fit a linear continuum using the two regions that border the CN emission band, which was then subtracted from the comet's spectra. We converted the band flux into column density using the $g$-factor from \citet{schleicher2010}, scaled to both the heliocentric distance and velocity (see Table \ref{observations}). To compute the gas production rate we assumed the Haser modelling with the outflow velocity $v_p$ scaled with $r_h$ ($v_p$ = 0.86 $r_h^{-4}$ km s$^{-1}$), customary values for the daughter velocity $v_d$ = 1 km s$^{-1}$, and scale lengths given in \citet{ahearn1995}. For the corresponding set of parameters in the Haser modelling, we produced theoretical column density profiles for CN varying the production rate until the best match between observations and theoretical predictions was achieved. We obtained $Q$(CN) = (2.3 $\pm$ 0.4) $\times$ 10$^{24}$ mol s$^{-1}$ and $Q$(CN) = (9.5 $\pm$ 0.2) $\times$ 10$^{24}$ mol s$^{-1}$ for the 2019, Sep. 24 and 26 nights, respectively. The first value is in good agreement with the set of measurements reported by \citet{opitom2019} from observations between 2019, Sep. 30 and October 13, ranging from $Q$(CN) = (1.6 $\pm$ 0.5) $\times$ 10$^{24}$ to $Q$(CN) = (2.1 $\pm$ 0.1) $\times$ 10$^{24}$ mol s$^{-1}$, while the second value is significantly larger than any of the reported values so far, including that from \citet{fitzsimmons2019}, $Q$(CN) = (3.7 $\pm$ 0.4) $\times$ 10$^{24}$ mol s$^{-1}$, and \citet{kareta2020}, $Q$(CN) = (5 $\pm$ 2) $\times$ 10$^{24}$ mol s$^{-1}$, both on the night of 2019, Sep. 20. This larger value of the gas production could be caused by an intrinsic variability in the comet's activity (an outburst), but in any case it has to be taken with caution. Although the seeing conditions were similar on the two nights, the comet was observed at a higher airmass (lower elevation) in the case of the second night (2019, Sep. 26). From the 2D spectra we can provide a 3 $\sigma$ upper limit for the C$_2$ production rate of $Q$(C$_2$) $<$ (4.5 $\pm$ 0.1) $\times$ 10$^{24}$ mol s$^{-1}$, in excellent agreement with the upper limit reported by \citet{fitzsimmons2019}. Our measured ratio upper limit of $Q$(CN)/$Q$(C$_2$) $<$ 0.511 is larger than the one reported by \citet{kareta2020}, $Q$(CN)/$Q$(C$_2$) $<$ 0.095, and also slightly larger than the value reported by \citet{opitom2019}, $Q$(CN)/$Q$(C$_2$) $<$ 0.3, but still lower than the value of 0.66 that marks the limit between carbon 'Depleted' comets and 'Typical' comets in the \citet{ahearn1995} taxonomy. In a very recent work, \citet{mckay2020} reported an H$_2$O production rate of (6.3 $\pm$ 1.5) $\times$ 10$^{26}$ mol s$^{-1}$ from observations using the 2.3m Astrophysical Research Consortium telescope, on the night of 2019, Oct. 11. This value was derived from the detection of [O$_\textrm{I}$] 6300 \AA \ emission line with ARCES high-resolution spectrograph. Using this H$_2$O production rate, we obtain a $Q$(CN)/$Q$(H$_2$O) ratio of 0.36 $\pm$ 0.17 per cent and 1.51 $\pm$ 0.26 per cent for the 2019, Sep. 24 and 26 nights, respectively. Again, our obtained $Q$(CN)/$Q$(H$_2$O) ratio for the 2019 Sep. 24 night is in good agreement with the values observed by \citet{ahearn1995} for comets in the Solar system as a function of heliocentric distance (see Fig. 2 in \citealt{mckay2020}), while the ratio obtained for the 2019, Sep. 26 night is larger than any other value obtained for comets at such heliocentric distance (2.58 au).

\subsection{Dynamics}
\label{dynamics}
   \subsubsection{Radiant}

The kinematic properties of minor bodies escaped from an Oort Cloud structure hosted by another star must be consistent with those of stars in the solar neighbourhood and beyond; therefore, the analysis of the pre-encounter trajectory of 2I/Borisov might shed some light on its most likely origin. 

We have performed integrations backward in time of 1000 control orbits of 2I/Borisov generated by applying the Monte Carlo using the Covariance Matrix (MCCM) method described by \citet{delafuente2015} and modified here to work with hyperbolic orbits. A statistical analysis (median and 16th and 84th percentiles) of the results indicates that at 1.650954 $\pm$ 0.000005~pc from the Sun and 5$\times$10$^{5}$~yr into the past, 2I/Borisov was moving inwards, at -32.28220$_{-0.00020}^{+0.00010}$~km~s$^{-1}$ with respect of the barycentre of the Solar system (1I/`Oumuamua had $-26$~km~s$^{-1}$, e.g. \citealt{mamajek2017}) and projected towards (radiant or antapex) $\alpha$ = 02$^{\rm h}$\,11$^{\rm m}$\,10$^{\rm s}$, $\delta$ = +59$\degr$\,26$\arcmin$\,24$\arcsec$ (32$\fdg$7894 $\pm$ 0.0002, +59$\fdg$44010 $\pm$ 0$\fdg$00010) in the constellation of Cassiopeia but close to Perseus as seen from Earth with Galactic coordinates $l$ = 132$\fdg$923, $b$ = -01$\fdg$876, and ecliptic coordinates $\lambda$ = 54$\fdg$321, $\beta$ = +42$\fdg$881. Therefore, the radiant appears projected against the plane of the Milky Way. The components of its Galactic velocity were $(U, V, W)$ = (22.00060 $\pm$ 0.00010, -23.60090 $\pm$ 0.00010, 1.056640$_{-0.00006}^{+0.000010})$ km s$^{-1}$. These values have been computed as described by \citet{delafuente2019_1}. It is rather difficult to find the exact origin of 2I/Borisov due to the limited availability of data for stars in the solar neighbourhood and beyond; however, its velocity matches well that of known neighbouring stars and also those of more distant regions of the Galactic disc. 

We have used {\it Gaia} second data release (DR2, \citealt{gaia2016,gaia2018}) to search for kinematic analogues of 2I/Borisov within 6000 $\sigma$ of the values of the Galactic velocity components presented above, as discussed by \citet{delafuente2019_2}. One of the relevant stars, TYC~3292-1633-1 (a somewhat solar-like star) is only about 90~pc from the Sun and it has $(U, V, W)$ = (22.5 $\pm$ 0.8, -23.4 $\pm$ 0.9, 1.3 $\pm$ 0.2) km s$^{-1}$. However, calculations carried out using the approach described in \citet{delafuente2018_1} failed to produce a consistently close flyby between star and comet. If more distant candidate sources are considered, the radiant appears projected towards the Perseus spiral arm about 2~kpc from the Sun, which is home of multiple regions of star formation. The Galactic velocity of 2I/Borisov prior to its encounter with the Solar system is compatible with that of the so-called Cassiopeia-Perseus family of open star clusters as described by \citet{delafuente2009}.
The origin of 2I/Borisov has been previously investigated by \citet{bailerjones2020} and \citet{hallatt2020}, both studies have been unable to single out robust extrasolar planetary system candidates to be the source of this interstellar comet.

   \subsubsection{Apex}
   Answering the question of where is 2I/Borisov headed next after its flyby with the Sun requires the analysis of a similar set of direct $N$-body simulations (also with MCCM), but forward in time. At 1.650668\,$\pm$\,0.000005~pc from the Sun and 5$\times$10$^{5}$~yr into the future, this object will be receding from us at 32.27590$_{-0.00020}^{+0.00010}$ km s$^{-1}$ towards (apex) $\alpha$ = 18$^{\rm h}$\,21$^{\rm m}$\,39$^{\rm s}$, $\delta$ = -51$\degr$\,58$\arcmin$\,37$\arcsec$ (275$\fdg$41350 $\pm$ 0\fdg00007, -51$\fdg$97680 $\pm$ 0$\fdg$00010) in the constellation of Telescopium but close to Ara with Galactic coordinates $l$ = 342$\fdg$563, $b$ = -16$\fdg$746, and ecliptic coordinates $\lambda$ = 273$\fdg$794, $\beta$ = -28$\fdg$608. Its post-encounter Galactic velocity will be (29.49770 $\pm$ 0.00010, -9.22751$_{-0.00008}^{+0.00005}$, -9.29887$_{-0.00005}^{+0.00006}$) km s$^{-1}$.

   \subsubsection{Non-gravitational forces}
The results presented above have been obtained using initial conditions derived from the orbit determination included in Table~\ref{elements} and performing integrations that neglected non-gravitational acceleration terms in the motion of 2I/Borisov. However, the orbit determination in Table~\ref{elements} is based on considering the vaporization of water ice on a rapidly rotating nucleus (Whipple icy model, see e.g \citealt{marsden1973}) with an active vent (see e.g \citealt{sekanina1993}) that produces a force with radial, transverse and normal components of the form $A\ g(r)$ with $g(r) = \alpha\ (r/r_0)^{-m} (1 + (r/r_0)^{n})^{-k}$, where $r_0$=2.808 au, $k$=4.6142, $m$=2.15, $n$=5.093, $\alpha$=0.1112620426, and the $A$-parameters have the values given in Table~\ref{elements}, with the normal component being of negligible value for this orbit determination. We have repeated the calculations discussed above including this H$_2$O-driven outgassing (full details of these calculations will be provided in a forthcoming publication), using again MCCM techniques to generate initial conditions.

 Including this non-gravitational force, the approach velocity of 2I/Borisov was -32.2815$_{-0.0003}^{+0.0004}$~km~s$^{-1}$ coming from coordinates $\alpha$ = 02$^{\rm h}$\,11$^{\rm m}$\,11$^{\rm s}$, $\delta$ = +59$\degr$\,26$\arcmin$\,27$\arcsec$ (32$\fdg$7936$_{-0\fdg0003}^{+0\fdg0002}$, +59$\fdg$44070 $\pm$ 0$\fdg${00010}); the components of its Galactic velocity were $(U, V, W)$ = (22.0009$_{-0.0003}^{+0.0002}$, -23.5996$_{-0.0002}^{+0.0003}$, 1.05592$_{-0.00007}^{+0.00003}$) km s$^{-1}$. Its future receding velocity will be 32.2759 $\pm$ 0.0002 km s$^{-1}$ moving towards $\alpha$ = 18$^{\rm h}$\,21$^{\rm m}$\,39$^{\rm s}$, $\delta$ = -51$\degr$\,58$\arcmin$\,37$\arcsec$ (275$\fdg$41350 $\pm$ 0$\fdg$00007, -51$\fdg$97680 $\pm$ 0$\fdg$00010); the components of its Galactic velocity were (29.49770 $\pm$ 0.00010, -9.22751 $\pm$ 0.00005, -9.29887$_{-0.00002}^{+0.00006}$) km s$^{-1}$. 

We have compared our results with predictions from JPL's SSDG SBDB and found discrepancies under 0.1 per cent with overlapping values within the computed uncertainties although the integration techniques and physical model used here are different. Therefore, we are confident that our results are sufficiently reliable. \citet{bailerjones2020} carried out a similar analysis using orbit determinations released prior to 2019 December and their overall conclusions are consistent with the ones presented here although the numerical values are slightly different.

\subsubsection{Kinematic context}

         Although all the calculations show that both 1I/`Oumuamua and 2I/Borisov arrived from interstellar space and they will return back to it, the kinematic signature of 2I/Borisov is different from that of 1I/`Oumuamua as described by e.g. \citet{mamajek2017}. 1I/`Oumuamua is apparently moving outwards, but 2I/Borisov is headed for the inner section of the Milky Way. The radiant is also different and far from others associated with known weakly hyperbolic comets (see e.g. fig.~3 in \citealt{delafuente2018_2}). Finding the actual origin of a given interstellar object is a difficult task now and it will continue being so for decades to come. The limiting factor is not in the uncertainty of the coordinates of the radiants or the values of the components of the Galactic velocity of the interstellar minor bodies, but in the quality and the quantity of the available stellar data. In general, Solar system data are more precise than their Galactic counterparts. In addition and for example, out of the billion of sources in {\it Gaia} DR2, less than eight millions have positions, parallax, radial velocity, and proper motions; these data are essential to compute the Galactic velocity of the stars, but also to generate initial conditions to perform $N$-body simulations to confirm possible close encounters at low relative velocity between interstellar objects and field stars.
         
\section{Conclusions}
\label{conclusions}

We have conducted an observational study of interstellar comet 2I/Borisov, using the 10.4-m GTC and the 3.6-m TNG telescopes, located at the Roque de los Muchachos Observatory, in the island of La Palma (Spain). We obtained a series of images in the visible using GTC and the Sloan-$r'$ filter, that allowed us to analyse Borisov's cometary activity and dust ejection. From the photometry of the images we inferred an upper limit for the size of the nucleus of $r_N <$ 3.6 km, a value that is in agreement with the one obtained from the dust modelling, and also in agreement with upper limits from other authors. In addition, dust modelling results indicate a moderate dust production rate of $\sim$ 50 kg s$^{-1}$ at heliocentric distance of $r_h$ = 2.6 au, with a differential power-law dust size distribution of index $\sim$ --3.4, in the range retrieved for many comet comae. Images in the near-infrared using $J$ and $H$ filters yielded a colour ($J - H$) = 0.52 $\pm$ 0.54, similar to the ($J - H$) colours found for comets and other related objects like TNOs and Centaurs in our Solar system. From the visible spectra obtained using the 10.4-m GTC telescope we computed a spectral slope of $S'$ = 12 $\pm$ 1 \%/1000\AA \ in the 4000 -- 9000\AA \ wavelength range, similar to the slope measured for 1I/`Oumuamua in the same wavelength range. The observed spectral gradient of comet 2I/Borisov, being redder at shorter wavelengths, is also in very good agreement with the observed spectral behaviour of cometary dust in Solar system comets. The visible spectrum allowed us to extract a colour ($g - r'$)  = 0.69 $\pm$ 0.04, which is comparable to the ($g - r'$) colour computed for 1I/Oumuamua, and very similar to the ($g - r'$) colours obtained for active and non-active comets in our Solar system. It also allowed us to measured CN gas production rates of $Q$(CN) = (2.3 $\pm$ 0.4) $\times$ 10$^{24}$ mol s$^{-1}$ and $Q$(CN) = (9.5 $\pm$ 0.2) $\times$ 10$^{24}$ mol s$^{-1}$, for the nights of 2019, Sep. 24 and Sep. 26, respectively. The former value is in good agreement with measurements reported by other authors on similar nights, while the latter is slightly larger than any reported CN gas production rate. A 3 $\sigma$ upper limit for the C$_2$ production rate of $Q$(C$_2$) $<$ (4.5 $\pm$ 0.1) $\times$ 10$^{24}$ mol s$^{-1}$ was obtained from the spectra, placing interstellar comet 2I/Borisov in the region of carbon 'Depleted' comets according to \citet{ahearn1995} taxonomy. Finally, our dynamical simulations showed that the Galactic velocity of 2I/Borisov matches well that of known stars in the solar neighbourhood and also those of more distant regions of the Galactic disc.

\section*{Acknowledgements}

JdL and MP acknowledge financial support from the project ProID2017010112 under the Operational Programmes of the European Regional Development Fund and the European Social Fund of the Canary Islands (OP-ERDF-ESF), as well as the Canarian Agency for Research, Innovation and Information Society (ACIISI), and the project AYA2017-89090-P of the Spanish Ministerio de Econom\'{\i}a y Competitividad' (MINECO). CdlFM and RdlFM thank S.~J. Aarseth for providing one of the codes used in this research and for comments on the implementation of non-gravitational forces in the calculations, and A.~I. G\'omez de Castro for providing access to computing facilities. Part of the calculations and the data analysis were completed on the Brigit HPC server of the `Universidad Complutense de Madrid' (UCM), and we thank S. Cano Als\'ua for his help during this stage. This work was partially supported by the Spanish MINECO under grant ESP2017-87813-R. In preparation of this paper, we made use of the NASA Astrophysics Data System, the ASTRO-PH e-print server, the MPC data server, and the SIMBAD and VizieR databases operated at CDS, Strasbourg, France. This work has made use of data from the European Space Agency (ESA) mission {\it Gaia} (\url{https://www.cosmos.esa.int/gaia}), processed by the {\it Gaia} Data Processing and Analysis Consortium (DPAC, \url{https://www.cosmos.esa.int/web/gaia/dpac/consortium}). Funding for the DPAC has been provided by national institutions, in particular the institutions participating in the {\it Gaia} Multilateral Agreement.. LML acknowledges the financial support from the State Agency for Research of the Spanish MCIU through the Centro de Excelencia Severo Ochoa Program under grant SEV-2017-0709, and from the research project PGC2018-099425-B-I00. FM acknowledges financial support from the Spanish Plan Nacional de Astronomia y Astrofisica LEONIDAS project RTI2018-095330-B-100 and the Centro de Excelencia Severo Ochoa Program under
grant SEV-2017-0709. NPA acknowledges funds through the SRI/FSI project ``Digging-Up Ice Rocks in the Solar System'' and the Center for Lunar and Asteroid Surface Science funded by NASA's SSERVI program at the University of Central Florida. MDP acknowledges funding from the Prominent Postdoctoral Program of the University of Central Florida. ACS-F acknowledges CAPES (Coordena\c c\~ao de Aperfei\c coamento de Pessoal de N\'ivel Superior - Brasil) for financed in part this study (Finance Code 001). Based on observations made with the Gran Telescopio Canarias (GTC) and the Italian Telescopio Nazionale Galileo (TNG), both installed at the Spanish Observatorio del  Roque de los Muchachos of the Instituto de Astrof\'isica de Canarias, in the island of La Palma. The TNG is operated by the Fundaci\'on Galileo Galilei of the INAF (Istituto Nazionale di Astrofisica). We thank E. Poretti for the allocation of Director's Discretionary Time at TNG.




\bibliographystyle{mnras}
\bibliography{BorisovReferences} 

\begin{thebibliography}{}
\makeatletter
\relax
\def\mn@urlcharsother{\let\do\@makeother \do\$\do\&\do\#\do\^\do\_\do\%\do\~}
\def\mn@doi{\begingroup\mn@urlcharsother \@ifnextchar [ {\mn@doi@}
  {\mn@doi@[]}}
\def\mn@doi@[#1]#2{\def\@tempa{#1}\ifx\@tempa\@empty \href
  {http://dx.doi.org/#2} {doi:#2}\else \href {http://dx.doi.org/#2} {#1}\fi
  \endgroup}
\def\mn@eprint#1#2{\mn@eprint@#1:#2::\@nil}
\def\mn@eprint@arXiv#1{\href {http://arxiv.org/abs/#1} {{\tt arXiv:#1}}}
\def\mn@eprint@dblp#1{\href {http://dblp.uni-trier.de/rec/bibtex/#1.xml}
  {dblp:#1}}
\def\mn@eprint@#1:#2:#3:#4\@nil{\def\@tempa {#1}\def\@tempb {#2}\def\@tempc
  {#3}\ifx \@tempc \@empty \let \@tempc \@tempb \let \@tempb \@tempa \fi \ifx
  \@tempb \@empty \def\@tempb {arXiv}\fi \@ifundefined
  {mn@eprint@\@tempb}{\@tempb:\@tempc}{\expandafter \expandafter \csname
  mn@eprint@\@tempb\endcsname \expandafter{\@tempc}}}

\bibitem[\protect\citeauthoryear{{A'Hearn}, {Millis}, {Schleicher}, {Osip}  \&
  {Birch}}{{A'Hearn} et~al.}{1995}]{ahearn1995}
{A'Hearn} M.~F.,  {Millis} R.~C.,  {Schleicher} D.~O.,  {Osip} D.~J.,   {Birch}
  P.~V.,  1995, \mn@doi [\icarus] {10.1006/icar.1995.1190}, \href
  {https://ui.adsabs.harvard.edu/abs/1995Icar..118..223A} {118, 223}

\bibitem[\protect\citeauthoryear{{Baffa} et~al.,}{{Baffa}
  et~al.}{2001}]{baffa2001}
{Baffa} C.,  et~al., 2001, \mn@doi [\aap] {10.1051/0004-6361:20011194}, \href
  {https://ui.adsabs.harvard.edu/abs/2001A&A...378..722B} {378, 722}

\bibitem[\protect\citeauthoryear{{Bailer-Jones}, {Farnocchia}, {Ye}, {Meech}
  \& {Micheli}}{{Bailer-Jones} et~al.}{2020}]{bailerjones2020}
{Bailer-Jones} C. A.~L.,  {Farnocchia} D.,  {Ye} Q.,  {Meech} K.~J.,
  {Micheli} M.,  2020, \mn@doi [\aap] {10.1051/0004-6361/201937231}, \href
  {https://ui.adsabs.harvard.edu/abs/2020A&A...634A..14B} {634, A14}

\bibitem[\protect\citeauthoryear{{Bannister} et~al.,}{{Bannister}
  et~al.}{2017}]{bannister2017}
{Bannister} M.~T.,  et~al., 2017, \mn@doi [\apjl] {10.3847/2041-8213/aaa07c},
  \href {https://ui.adsabs.harvard.edu/abs/2017ApJ...851L..38B} {851, L38}

\bibitem[\protect\citeauthoryear{{Boe} et~al.,}{{Boe} et~al.}{2019}]{boe2019}
{Boe} B.,  et~al., 2019, \mn@doi [\icarus] {10.1016/j.icarus.2019.05.034},
  \href {https://ui.adsabs.harvard.edu/abs/2019Icar..333..252B} {333, 252}

\bibitem[\protect\citeauthoryear{{Bohlin}, {Gordon}  \& {Tremblay}}{{Bohlin}
  et~al.}{2014}]{bohlin2014}
{Bohlin} R.~C.,  {Gordon} K.~D.,   {Tremblay} P.~E.,  2014, \mn@doi [\pasp]
  {10.1086/677655}, \href
  {https://ui.adsabs.harvard.edu/abs/2014PASP..126..711B} {126, 711}

\bibitem[\protect\citeauthoryear{{Bolin} et~al.,}{{Bolin}
  et~al.}{2018}]{bolin2018}
{Bolin} B.~T.,  et~al., 2018, \mn@doi [\apjl] {10.3847/2041-8213/aaa0c9}, \href
  {https://ui.adsabs.harvard.edu/abs/2018ApJ...852L...2B} {852, L2}

\bibitem[\protect\citeauthoryear{{Cepa}}{{Cepa}}{2010}]{cepa2010}
{Cepa} J.,  2010, \mn@doi [Astrophysics and Space Science Proceedings]
  {10.1007/978-3-642-11250-8_2}, \href
  {https://ui.adsabs.harvard.edu/abs/2010ASSP...14...15C} {14, 15}

\bibitem[\protect\citeauthoryear{{Cepa} et~al.,}{{Cepa}
  et~al.}{2000}]{cepa2000}
{Cepa} J.,  et~al., 2000, in {Iye} M.,  {Moorwood} A.~F.,  eds,  Society of
  Photo-Optical Instrumentation Engineers (SPIE) Conference Series Vol. 4008,
  \procspie. pp 623--631, \mn@doi{10.1117/12.395520}

\bibitem[\protect\citeauthoryear{{Cook}, {Ragozzine}, {Granvik}  \&
  {Stephens}}{{Cook} et~al.}{2016}]{cook2016}
{Cook} N.~V.,  {Ragozzine} D.,  {Granvik} M.,   {Stephens} D.~C.,  2016,
  \mn@doi [\apj] {10.3847/0004-637X/825/1/51}, \href
  {https://ui.adsabs.harvard.edu/abs/2016ApJ...825...51C} {825, 51}

\bibitem[\protect\citeauthoryear{{de la Fuente Marcos} \& {de la Fuente
  Marcos}}{{de la Fuente Marcos} \& {de la Fuente
  Marcos}}{2009}]{delafuente2009}
{de la Fuente Marcos} R.,  {de la Fuente Marcos} C.,  2009, \mn@doi [\na]
  {10.1016/j.newast.2008.08.001}, \href
  {https://ui.adsabs.harvard.edu/abs/2009NewA...14..180D} {14, 180}

\bibitem[\protect\citeauthoryear{{de la Fuente Marcos} \& {de la Fuente
  Marcos}}{{de la Fuente Marcos} \& {de la Fuente
  Marcos}}{2015}]{delafuente2015}
{de la Fuente Marcos} C.,  {de la Fuente Marcos} R.,  2015, \mn@doi [\mnras]
  {10.1093/mnras/stv1725}, \href
  {https://ui.adsabs.harvard.edu/abs/2015MNRAS.453.1288D} {453, 1288}

\bibitem[\protect\citeauthoryear{{de la Fuente Marcos} \& {de la Fuente
  Marcos}}{{de la Fuente Marcos} \& {de la Fuente
  Marcos}}{2018}]{delafuente2018_1}
{de la Fuente Marcos} R.,  {de la Fuente Marcos} C.,  2018, \mn@doi [Research
  Notes of the American Astronomical Society] {10.3847/2515-5172/aac2d0}, \href
  {https://ui.adsabs.harvard.edu/abs/2018RNAAS...2...30D} {2, 30}

\bibitem[\protect\citeauthoryear{{de la Fuente Marcos} \& {de la Fuente
  Marcos}}{{de la Fuente Marcos} \& {de la Fuente
  Marcos}}{2019a}]{delafuente2019_2}
{de la Fuente Marcos} C.,  {de la Fuente Marcos} R.,  2019a, \mn@doi [\mnras]
  {10.1093/mnras/stz2229}, \href
  {https://ui.adsabs.harvard.edu/abs/2019MNRAS.489..951D} {489, 951}

\bibitem[\protect\citeauthoryear{{de la Fuente Marcos} \& {de la Fuente
  Marcos}}{{de la Fuente Marcos} \& {de la Fuente
  Marcos}}{2019b}]{delafuente2019_1}
{de la Fuente Marcos} R.,  {de la Fuente Marcos} C.,  2019b, \mn@doi [\aap]
  {10.1051/0004-6361/201935008}, \href
  {https://ui.adsabs.harvard.edu/abs/2019A&A...627A.104D} {627, A104}

\bibitem[\protect\citeauthoryear{{de la Fuente Marcos}, {de la Fuente Marcos}
  \& {Aarseth}}{{de la Fuente Marcos} et~al.}{2018}]{delafuente2018_2}
{de la Fuente Marcos} C.,  {de la Fuente Marcos} R.,   {Aarseth} S.~J.,  2018,
  \mn@doi [\mnras] {10.1093/mnrasl/sly019}, \href
  {https://ui.adsabs.harvard.edu/abs/2018MNRAS.476L...1D} {476, L1}

\bibitem[\protect\citeauthoryear{de Le{\'{o}}n, Licandro, Serra-Ricart,
  Cabrera-Lavers, Serra, Scarpa, de~la Fuente~Marcos  \& de~la
  Fuente~Marcos}{de~Le{\'{o}}n et~al.}{2019}]{deleon2019}
de Le{\'{o}}n J.,  Licandro J.,  Serra-Ricart M.,  Cabrera-Lavers A.,  Serra
  J.~F.,  Scarpa R.,  de~la Fuente~Marcos C.,   de~la Fuente~Marcos R.,  2019,
  \mn@doi [Research Notes of the {AAS}] {10.3847/2515-5172/ab449c}, 3, 131

\bibitem[\protect\citeauthoryear{{Della Corte} et~al.,}{{Della Corte}
  et~al.}{2015}]{dellacorte2015}
{Della Corte} V.,  et~al., 2015, \mn@doi [\aap] {10.1051/0004-6361/201526208},
  \href {https://ui.adsabs.harvard.edu/abs/2015A&A...583A..13D} {583, A13}

\bibitem[\protect\citeauthoryear{{Della Corte} et~al.,}{{Della Corte}
  et~al.}{2016}]{dellacorte2016}
{Della Corte} V.,  et~al., 2016, \mn@doi [\mnras] {10.1093/mnras/stw2529},
  \href {https://ui.adsabs.harvard.edu/abs/2016MNRAS.462S.210D} {462, S210}

\bibitem[\protect\citeauthoryear{{Engelhardt}, {Vere{\v{s}}}, {Jedicke},
  {Denneau}  \& {Beshore}}{{Engelhardt} et~al.}{2014}]{engelhardt2014}
{Engelhardt} T.,  {Vere{\v{s}}} P.,  {Jedicke} R.,  {Denneau} L.,   {Beshore}
  E.,  2014, in {Muinonen} K.,  {Penttil{\"a}} A.,  {Granvik} M.,  {Virkki} A.,
   {Fedorets} G.,  {Wilkman} O.,   {Kohout} T.,  eds, Asteroids, Comets,
  Meteors 2014. p.~149

\bibitem[\protect\citeauthoryear{{Fitzsimmons} et~al.,}{{Fitzsimmons}
  et~al.}{2018}]{fitzsimmons2018}
{Fitzsimmons} A.,  et~al., 2018, \mn@doi [Nature Astronomy]
  {10.1038/s41550-017-0361-4}, \href
  {https://ui.adsabs.harvard.edu/abs/2018NatAs...2..133F} {2, 133}

\bibitem[\protect\citeauthoryear{{Fitzsimmons} et~al.,}{{Fitzsimmons}
  et~al.}{2019}]{fitzsimmons2019}
{Fitzsimmons} A.,  et~al., 2019, \mn@doi [\apjl] {10.3847/2041-8213/ab49fc},
  \href {https://ui.adsabs.harvard.edu/abs/2019ApJ...885L...9F} {885, L9}

\bibitem[\protect\citeauthoryear{{Fulle}}{{Fulle}}{2004}]{fulle2004}
{Fulle} M.,  2004, in {Festou} M.~C.,  {Keller} H.~U.,   {Weaver} H.~A.,  eds,
  Comets II. University of Arizona Press, p.~565

\bibitem[\protect\citeauthoryear{{Fulle} et~al.,}{{Fulle}
  et~al.}{2016}]{fulle2016}
{Fulle} M.,  et~al., 2016, \mn@doi [\apj] {10.3847/0004-637X/821/1/19}, \href
  {https://ui.adsabs.harvard.edu/abs/2016ApJ...821...19F} {821, 19}

\bibitem[\protect\citeauthoryear{{Gaia Collaboration} et~al.,}{{Gaia
  Collaboration} et~al.}{2016}]{gaia2016}
{Gaia Collaboration} et~al., 2016, \mn@doi [\aap]
  {10.1051/0004-6361/201629272}, \href
  {https://ui.adsabs.harvard.edu/abs/2016A&A...595A...1G} {595, A1}

\bibitem[\protect\citeauthoryear{{Gaia Collaboration} et~al.,}{{Gaia
  Collaboration} et~al.}{2018}]{gaia2018}
{Gaia Collaboration} et~al., 2018, \mn@doi [\aap]
  {10.1051/0004-6361/201833051}, \href
  {https://ui.adsabs.harvard.edu/abs/2018A&A...616A...1G} {616, A1}

\bibitem[\protect\citeauthoryear{{Guzik}, {Drahus}, {Rusek}, {Waniak},
  {Cannizzaro}  \& {Pastor-Marazuela}}{{Guzik} et~al.}{2020}]{guzik2020}
{Guzik} P.,  {Drahus} M.,  {Rusek} K.,  {Waniak} W.,  {Cannizzaro} G.,
  {Pastor-Marazuela} I.,  2020, \mn@doi [Nature Astronomy]
  {10.1038/s41550-019-0931-8}, \href
  {https://ui.adsabs.harvard.edu/abs/2020NatAs...4...53G} {4, 53}

\bibitem[\protect\citeauthoryear{{Hainaut}, {Boehnhardt}  \&
  {Protopapa}}{{Hainaut} et~al.}{2012}]{hainaut2012}
{Hainaut} O.~R.,  {Boehnhardt} H.,   {Protopapa} S.,  2012, \mn@doi [\aap]
  {10.1051/0004-6361/201219566}, \href
  {https://ui.adsabs.harvard.edu/abs/2012A&A...546A.115H} {546, A115}

\bibitem[\protect\citeauthoryear{{Hallatt} \& {Wiegert}}{{Hallatt} \&
  {Wiegert}}{2020}]{hallatt2020}
{Hallatt} T.,  {Wiegert} P.,  2020, \mn@doi [\aj] {10.3847/1538-3881/ab7336},
  \href {https://ui.adsabs.harvard.edu/abs/2020AJ....159..147H} {159, 147}

\bibitem[\protect\citeauthoryear{{Hanner}, {Tokunaga}, {Veeder}  \&
  {A'Hearn}}{{Hanner} et~al.}{1984}]{hanner1984}
{Hanner} M.~S.,  {Tokunaga} A.~T.,  {Veeder} G.~J.,   {A'Hearn} M.~F.,  1984,
  \mn@doi [\aj] {10.1086/113495}, \href
  {https://ui.adsabs.harvard.edu/abs/1984AJ.....89..162H} {89, 162}

\bibitem[\protect\citeauthoryear{{Hansen} \& {Travis}}{{Hansen} \&
  {Travis}}{1974}]{hansen1974}
{Hansen} J.~E.,  {Travis} L.~D.,  1974, \mn@doi [\ssr] {10.1007/BF00168069},
  \href {https://ui.adsabs.harvard.edu/abs/1974SSRv...16..527H} {16, 527}

\bibitem[\protect\citeauthoryear{{Holmberg}, {Flynn}  \&
  {Portinari}}{{Holmberg} et~al.}{2006}]{holmberg2006}
{Holmberg} J.,  {Flynn} C.,   {Portinari} L.,  2006, \mn@doi [\mnras]
  {10.1111/j.1365-2966.2005.09832.x}, \href
  {https://ui.adsabs.harvard.edu/abs/2006MNRAS.367..449H} {367, 449}

\bibitem[\protect\citeauthoryear{{Jester} et~al.,}{{Jester}
  et~al.}{2005}]{jester2005}
{Jester} S.,  et~al., 2005, \mn@doi [\aj] {10.1086/432466}, \href
  {https://ui.adsabs.harvard.edu/abs/2005AJ....130..873J} {130, 873}

\bibitem[\protect\citeauthoryear{{Jewitt}}{{Jewitt}}{2015}]{jewitt2015}
{Jewitt} D.,  2015, \mn@doi [\aj] {10.1088/0004-6256/150/6/201}, \href
  {https://ui.adsabs.harvard.edu/abs/2015AJ....150..201J} {150, 201}

\bibitem[\protect\citeauthoryear{{Jewitt} \& {Luu}}{{Jewitt} \&
  {Luu}}{2019}]{jewitt2019}
{Jewitt} D.,  {Luu} J.,  2019, \mn@doi [\apjl] {10.3847/2041-8213/ab530b},
  \href {https://ui.adsabs.harvard.edu/abs/2019ApJ...886L..29J} {886, L29}

\bibitem[\protect\citeauthoryear{{Jewitt}, {Luu}, {Rajagopal}, {Kotulla},
  {Ridgway}, {Liu}  \& {Augusteijn}}{{Jewitt} et~al.}{2017}]{jewitt2017}
{Jewitt} D.,  {Luu} J.,  {Rajagopal} J.,  {Kotulla} R.,  {Ridgway} S.,  {Liu}
  W.,   {Augusteijn} T.,  2017, \mn@doi [\apjl] {10.3847/2041-8213/aa9b2f},
  \href {https://ui.adsabs.harvard.edu/abs/2017ApJ...850L..36J} {850, L36}

\bibitem[\protect\citeauthoryear{{Jewitt}, {Hui}, {Kim}, {Mutchler}, {Weaver}
  \& {Agarwal}}{{Jewitt} et~al.}{2020}]{jewitt2020}
{Jewitt} D.,  {Hui} M.-T.,  {Kim} Y.,  {Mutchler} M.,  {Weaver} H.,   {Agarwal}
  J.,  2020, \mn@doi [\apjl] {10.3847/2041-8213/ab621b}, \href
  {https://ui.adsabs.harvard.edu/abs/2020ApJ...888L..23J} {888, L23}

\bibitem[\protect\citeauthoryear{{Kareta} et~al.,}{{Kareta}
  et~al.}{2020}]{kareta2020}
{Kareta} T.,  et~al., 2020, \mn@doi [\apjl] {10.3847/2041-8213/ab6a08}, \href
  {https://ui.adsabs.harvard.edu/abs/2020ApJ...889L..38K} {889, L38}

\bibitem[\protect\citeauthoryear{{Kochergin} et~al.,}{{Kochergin}
  et~al.}{2019}]{kochergin2019}
{Kochergin} A.,  et~al., 2019, \mn@doi [Research Notes of the American
  Astronomical Society] {10.3847/2515-5172/ab4c46}, \href
  {https://ui.adsabs.harvard.edu/abs/2019RNAAS...3..152K} {3, 152}

\bibitem[\protect\citeauthoryear{{Lamy} \& {Toth}}{{Lamy} \&
  {Toth}}{2009}]{lamy2009}
{Lamy} P.,  {Toth} I.,  2009, \mn@doi [\icarus] {10.1016/j.icarus.2009.01.030},
  \href {https://ui.adsabs.harvard.edu/abs/2009Icar..201..674L} {201, 674}

\bibitem[\protect\citeauthoryear{{Landolt}}{{Landolt}}{1992}]{landolt1992}
{Landolt} A.~U.,  1992, \mn@doi [\aj] {10.1086/116242}, \href
  {https://ui.adsabs.harvard.edu/abs/1992AJ....104..340L} {104, 340}

\bibitem[\protect\citeauthoryear{{Lee}, {Lin}, {Chen}  \& {Yen}}{{Lee}
  et~al.}{2019}]{lee2019}
{Lee} C.-H.,  {Lin} H.-W.,  {Chen} Y.-T.,   {Yen} S.-F.,  2019, \mn@doi
  [Research Notes of the American Astronomical Society]
  {10.3847/2515-5172/ab5f69}, \href
  {https://ui.adsabs.harvard.edu/abs/2019RNAAS...3..184L} {3, 184}

\bibitem[\protect\citeauthoryear{{Licandro}, {Popescu}, {de Le{\'o}n},
  {Morate}, {Vaduvescu}, {De Pr{\'a}}  \& {Ali-Laoga}}{{Licandro}
  et~al.}{2018}]{licandro2018}
{Licandro} J.,  {Popescu} M.,  {de Le{\'o}n} J.,  {Morate} D.,  {Vaduvescu} O.,
   {De Pr{\'a}} M.,   {Ali-Laoga} V.,  2018, \mn@doi [\aap]
  {10.1051/0004-6361/201832853}, \href
  {https://ui.adsabs.harvard.edu/abs/2018A&A...618A.170L} {618, A170}

\bibitem[\protect\citeauthoryear{{Licandro}, {de la Fuente Marcos}, {de la
  Fuente Marcos}, {de Le{\'o}n}, {Serra-Ricart}  \&
  {Cabrera-Lavers}}{{Licandro} et~al.}{2019}]{licandro2019}
{Licandro} J.,  {de la Fuente Marcos} C.,  {de la Fuente Marcos} R.,  {de
  Le{\'o}n} J.,  {Serra-Ricart} M.,   {Cabrera-Lavers} A.,  2019, \mn@doi
  [\aap] {10.1051/0004-6361/201834902}, \href
  {https://ui.adsabs.harvard.edu/abs/2019A&A...625A.133L} {625, A133}

\bibitem[\protect\citeauthoryear{{Luu} \& {Jewitt}}{{Luu} \&
  {Jewitt}}{1996}]{luu1996}
{Luu} J.~X.,  {Jewitt} D.~C.,  1996, \mn@doi [\aj] {10.1086/117801}, \href
  {https://ui.adsabs.harvard.edu/abs/1996AJ....111..499L} {111, 499}

\bibitem[\protect\citeauthoryear{{Mamajek}}{{Mamajek}}{2017}]{mamajek2017}
{Mamajek} E.,  2017, \mn@doi [Research Notes of the American Astronomical
  Society] {10.3847/2515-5172/aa9bdc}, \href
  {https://ui.adsabs.harvard.edu/abs/2017RNAAS...1...21M} {1, 21}

\bibitem[\protect\citeauthoryear{{Marsden}, {Sekanina}  \& {Yeomans}}{{Marsden}
  et~al.}{1973}]{marsden1973}
{Marsden} B.~G.,  {Sekanina} Z.,   {Yeomans} D.~K.,  1973, \mn@doi [\aj]
  {10.1086/111402}, \href
  {https://ui.adsabs.harvard.edu/abs/1973AJ.....78..211M} {78, 211}

\bibitem[\protect\citeauthoryear{{McGlynn} \& {Chapman}}{{McGlynn} \&
  {Chapman}}{1989}]{mcglynn1989}
{McGlynn} T.~A.,  {Chapman} R.~D.,  1989, \mn@doi [\apjl] {10.1086/185590},
  \href {https://ui.adsabs.harvard.edu/abs/1989ApJ...346L.105M} {346, L105}

\bibitem[\protect\citeauthoryear{{McKay}, {Cochran}, {Dello Russo}  \&
  {DiSanti}}{{McKay} et~al.}{2020}]{mckay2020}
{McKay} A.~J.,  {Cochran} A.~L.,  {Dello Russo} N.,   {DiSanti} M.~A.,  2020,
  \mn@doi [\apjl] {10.3847/2041-8213/ab64ed}, \href
  {https://ui.adsabs.harvard.edu/abs/2020ApJ...889L..10M} {889, L10}

\bibitem[\protect\citeauthoryear{Meech}{Meech}{2017}]{meech2017}
Meech K.~J.,  2017, \mn@doi [Philosophical Transactions of the Royal Society A:
  Mathematical, Physical and Engineering Sciences] {10.1098/rsta.2016.0247},
  375, 20160247

\bibitem[\protect\citeauthoryear{{Micheli} et~al.,}{{Micheli}
  et~al.}{2018}]{micheli2018}
{Micheli} M.,  et~al., 2018, \mn@doi [\nat] {10.1038/s41586-018-0254-4}, \href
  {https://ui.adsabs.harvard.edu/abs/2018Natur.559..223M} {559, 223}

\bibitem[\protect\citeauthoryear{{Moreno} et~al.,}{{Moreno}
  et~al.}{2016}]{moreno2016}
{Moreno} F.,  et~al., 2016, \mn@doi [\aap] {10.1051/0004-6361/201527564}, \href
  {https://ui.adsabs.harvard.edu/abs/2016A&A...587A.155M} {587, A155}

\bibitem[\protect\citeauthoryear{{Moreno} et~al.,}{{Moreno}
  et~al.}{2017}]{moreno2017}
{Moreno} F.,  et~al., 2017, \mn@doi [\apjl] {10.3847/2041-8213/aa6036}, \href
  {https://ui.adsabs.harvard.edu/abs/2017ApJ...837L...3M} {837, L3}

\bibitem[\protect\citeauthoryear{Nelder \& Mead}{Nelder \&
  Mead}{1965}]{nelder1965}
Nelder J.~A.,  Mead R.,  1965, \mn@doi [The Computer Journal]
  {10.1093/comjnl/7.4.308}, 7, 308

\bibitem[\protect\citeauthoryear{{Opitom} et~al.,}{{Opitom}
  et~al.}{2019}]{opitom2019}
{Opitom} C.,  et~al., 2019, \mn@doi [\aap] {10.1051/0004-6361/201936959}, \href
  {https://ui.adsabs.harvard.edu/abs/2019A&A...631L...8O} {631, L8}

\bibitem[\protect\citeauthoryear{{'Oumuamua ISSI Team} et~al.,}{{'Oumuamua ISSI
  Team} et~al.}{2019}]{bannister2019}
{'Oumuamua ISSI Team} et~al., 2019, \mn@doi [Nature Astronomy]
  {10.1038/s41550-019-0816-x}, \href
  {https://ui.adsabs.harvard.edu/abs/2019NatAs...3..594O} {3, 594}

\bibitem[\protect\citeauthoryear{{Persson}, {Murphy}, {Krzeminski}, {Roth}  \&
  {Rieke}}{{Persson} et~al.}{1998}]{persson1998}
{Persson} S.~E.,  {Murphy} D.~C.,  {Krzeminski} W.,  {Roth} M.,   {Rieke}
  M.~J.,  1998, \mn@doi [\aj] {10.1086/300607}, \href
  {https://ui.adsabs.harvard.edu/abs/1998AJ....116.2475P} {116, 2475}

\bibitem[\protect\citeauthoryear{{Picazzio}, {Figueredo}, {de Almeida}, {de
  Oliveira}  \& {Churyumov}}{{Picazzio} et~al.}{2010}]{picazzio2010}
{Picazzio} E.,  {Figueredo} E.,  {de Almeida} A.~A.,  {de Oliveira} C.~M.,
  {Churyumov} K.~I.,  2010, in {Fernandez} J.~A.,  {Lazzaro} D.,  {Prialnik}
  D.,   {Schulz} R.,  eds,  IAU Symposium Vol. 263, Icy Bodies of the Solar
  System. pp 285--288, \mn@doi{10.1017/S1743921310001948}

\bibitem[\protect\citeauthoryear{{Popescu} et~al.,}{{Popescu}
  et~al.}{2016}]{popescu2016}
{Popescu} M.,  et~al., 2016, \mn@doi [\aap] {10.1051/0004-6361/201628163},
  \href {https://ui.adsabs.harvard.edu/abs/2016A&A...591A.115P} {591, A115}

\bibitem[\protect\citeauthoryear{{Rotundi} et~al.,}{{Rotundi}
  et~al.}{2015}]{rotundi2015}
{Rotundi} A.,  et~al., 2015, \mn@doi [Science] {10.1126/science.aaa3905}, \href
  {https://ui.adsabs.harvard.edu/abs/2015Sci...347a3905R} {347, aaa3905}

\bibitem[\protect\citeauthoryear{{Schleicher}}{{Schleicher}}{2010}]{schleicher2010}
{Schleicher} D.~G.,  2010, \mn@doi [\aj] {10.1088/0004-6256/140/4/973}, \href
  {https://ui.adsabs.harvard.edu/abs/2010AJ....140..973S} {140, 973}

\bibitem[\protect\citeauthoryear{{Sekanina}}{{Sekanina}}{1993}]{sekanina1993}
{Sekanina} Z.,  1993, \aap, \href
  {https://ui.adsabs.harvard.edu/abs/1993A&A...277..265S} {277, 265}

\bibitem[\protect\citeauthoryear{{Sen} \& {Rana}}{{Sen} \&
  {Rana}}{1993}]{sen1993}
{Sen} A.~K.,  {Rana} N.~C.,  1993, \aap, \href
  {https://ui.adsabs.harvard.edu/abs/1993A&A...275..298S} {275, 298}

\bibitem[\protect\citeauthoryear{{Solontoi} et~al.,}{{Solontoi}
  et~al.}{2012}]{solontoi2012}
{Solontoi} M.,  et~al., 2012, \mn@doi [\icarus] {10.1016/j.icarus.2011.10.008},
  \href {https://ui.adsabs.harvard.edu/abs/2012Icar..218..571S} {218, 571}

\bibitem[\protect\citeauthoryear{{Stern}}{{Stern}}{1990}]{stern1990}
{Stern} S.~A.,  1990, \mn@doi [\pasp] {10.1086/132704}, \href
  {https://ui.adsabs.harvard.edu/abs/1990PASP..102..793S} {102, 793}

\bibitem[\protect\citeauthoryear{{Sykes}, {Cutri}, {Fowler}, {Tholen},
  {Skrutskie}, {Price}  \& {Tedesco}}{{Sykes} et~al.}{2000}]{sykes2000}
{Sykes} M.~V.,  {Cutri} R.~M.,  {Fowler} J.~W.,  {Tholen} D.~J.,  {Skrutskie}
  M.~F.,  {Price} S.,   {Tedesco} E.~F.,  2000, \mn@doi [\icarus]
  {10.1006/icar.2000.6366}, \href
  {https://ui.adsabs.harvard.edu/abs/2000Icar..146..161S} {146, 161}

\bibitem[\protect\citeauthoryear{{Ye}, {Zhang}, {Kelley}  \& {Brown}}{{Ye}
  et~al.}{2017}]{ye2017}
{Ye} Q.-Z.,  {Zhang} Q.,  {Kelley} M. S.~P.,   {Brown} P.~G.,  2017, \mn@doi
  [\apjl] {10.3847/2041-8213/aa9a34}, \href
  {https://ui.adsabs.harvard.edu/abs/2017ApJ...851L...5Y} {851, L5}

\bibitem[\protect\citeauthoryear{{Ye} et~al.,}{{Ye} et~al.}{2020}]{ye2020}
{Ye} Q.,  et~al., 2020, \mn@doi [\aj] {10.3847/1538-3881/ab659b}, \href
  {https://ui.adsabs.harvard.edu/abs/2020AJ....159...77Y} {159, 77}

\makeatother
\end{thebibliography}








\bsp	
\label{lastpage}
\end{document}